\documentclass[preprint,12pt,3p,sort&compress]{elsarticle}

\usepackage{csquotes}
\usepackage{caption}
\usepackage{ upgreek }
\usepackage{subcaption}
\usepackage{easyReview}
\usepackage{natbib}
\usepackage{amssymb,bm}
\usepackage[mathcal]{euscript}
\usepackage{epstopdf}
\usepackage{amsthm,amsmath,mathtools}

\usepackage{color}
\usepackage{mathrsfs}
\usepackage{dsfont}
\usepackage{graphicx}
\usepackage[shortlabels]{enumitem}
\usepackage{empheq}
\usepackage{float}
\usepackage{physics}
\usepackage[linkcolor=blue, urlcolor=blue, citecolor=blue,
colorlinks, bookmarks]{hyperref}
\usepackage{amsfonts}
\usepackage{multirow,booktabs}
\usepackage{adjustbox}
\usepackage{cases} 
\usepackage{algorithm}       
\usepackage{algpseudocode}
\numberwithin{equation}{section}

\numberwithin{Example}{section}
\usepackage{array}
\newcolumntype{L}{>{$}l<{$}} 
\newcolumntype{C}{>{$}c<{$}}

\begin{document}	
	
	\begin{frontmatter}
	\title{Adaptive Sampling for Hydrodynamic Stability }
		\author{Anshima Singh}
		 \ead{anshima.singh@manchester.ac.uk}
             \author{David J. Silvester}
            \address{Department of Mathematics, University of Manchester, Oxford Road, Manchester M13 9PL, UK}
 \ead{david.silvester@manchester.ac.uk}
 
\begin{abstract}
An adaptive sampling approach for efficient detection of bifurcation boundaries in parametrized fluid flow problems is presented herein. The study extends the machine-learning approach of Silvester~(J. Comput. Phys., 553 (2026), 114743), where a classifier network was trained on preselected simulation data to identify bifurcated and nonbifurcated flow regimes. In contrast, the proposed methodology introduces adaptivity through a flow-based deep generative model that automatically refines the sampling of the parameter space. The strategy has two components: a classifier network maps the flow parameters to a bifurcation probability, and a probability density estimation technique (KRnet) for the generation of new samples at each adaptive step. The classifier output provides a probabilistic measure of flow stability, and the Shannon entropy of these predictions is employed as an uncertainty indicator. KRnet is trained to approximate a probability density function that concentrates sampling in regions of high entropy, thereby directing computational effort towards the evolving bifurcation boundary. This coupling between classification and generative modeling establishes a feedback-driven adaptive learning process analogous to error-indicator based refinement in contemporary partial differential equation solution strategies. Starting from a uniform parameter distribution, the new approach achieves accurate bifurcation boundary identification with significantly fewer Navier--Stokes simulations, providing a scalable foundation for high-dimensional stability analysis.
\end{abstract}

\end{frontmatter}

\section{Introduction}\label{sec:Intro}

The study of hydrodynamic stability has a long history, forming one of the central pillars of fluid mechanics and applied mathematics (see~\cite{joseph2013stability, landau1987fluid, MR604359, MR1801992}). Classical linear stability theory studies the evolution of small perturbations to a steady base flow and determines whether such disturbances decay or amplify over time (see~\cite{MR604359, MR1801992, MR1883627}). When perturbations grow, their amplification can be indicative of a loss of stability, often manifested as a bifurcation marking the transition away from the underlying steady or symmetric state.
This theory proceeds by linearising the governing Navier--Stokes equations about a reference solution and studying an associated eigenvalue problem, which provides a computational framework for tracking the critical eigenvalues that mark the onset of instability (see~\cite{MR1883627, Wubs2023}).
However, linear theory provides information only about the response to infinitesimal disturbances and therefore cannot quantify the role of finite-amplitude perturbations in the promotion of instability. Specifically, eigenvalue analysis offers only local insight into stability and does not account for the transient growth of finite perturbations due to non-normality of the associated eigenfunctions. These limitations motivate the exploration of data-driven approaches for illustrating flow stability and mapping bifurcation behaviour.

A growing body of work has explored machine-learning methods for analysing dynamical systems and detecting bifurcation behaviour. Several recent studies have shown that data-driven models can efficiently identify multiple bifurcations in low-dimensional systems~\cite{szep2021parameter, mogharabin2025bifurcation}. Related approaches have employed machine-learning techniques to classify trajectories as regular or chaotic from time-series data, and to approximate Lyapunov exponents using neural-network-based surrogates~\cite{celletti2022classification, MR4844871}. In the context of fluid mechanics, physics-informed neural networks have attracted significant attention following the pioneering work of Raissi et al.~\cite{MR4265157}, although subsequent studies have highlighted the challenges of applying PINNs to the Navier--Stokes equations~\cite{wang2023expert}. Other machine-learning strategies for steady-state flow bifurcations rely on reduced-order modelling approaches~\cite{MR4547235}, which require extensive snapshot datasets and introduce truncation errors that are difficult to quantify. 

More recently, the results in~\cite{silvester2026machine} demonstrate that even relatively simple neural network classifiers can accurately distinguish between bifurcated and nonbifurcated flow states when trained on sufficiently informative data, supporting the view that supervised learning can serve as an effective alternative to classical eigenvalue-based stability analysis. One issue, however, is that the training sets in this work were assembled manually using dense parameter sweeps informed by prior knowledge of where the transition was expected to occur. Since each labelled sample required a full computational fluid dynamics (CFD) simulation the generation of a reliable dataset demanded a substantial number of computationally expensive flow solves. This limitation motivates the development of sampling strategies that automatically concentrate computational effort in the most informative regions of the parameter domain.

In this work, we address this challenge by extending the supervised bifurcation classification approach to a fully adaptive, data-efficient methodology. Whereas the earlier study relied on a fixed, manually assembled training set, the present work introduces an adaptive sampling mechanism that dynamically concentrates new simulations where they are most informative. Our approach couples a neural-network classifier with a normalising-flow-based generative model (KRnet)~\cite{tang2020deep,MR4385873} that identifies regions of high predictive uncertainty and selectively proposes new parameter points where additional CFD simulations are most valuable. This adaptive refinement process progressively enriches the training set around the evolving stability boundary, improving classifier accuracy while substantially reducing the number of high-fidelity simulations required. The resulting method automates the exploration of nonlinear bifurcation structures in multidimensional parameter domains without relying on eigenvalue computations, or dense parameter sweeps.

In all problems considered in this study, the labelled flow states used to train the classifier are obtained from high-fidelity simulations performed with the IFISS software package~\cite{silvester2023ifiss}. Each solution is categorised as bifurcated or nonbifurcated using the kinetic-energy criterion introduced in~\cite{silvester2026machine}, which has been shown to provide a robust and consistent labelling strategy. This ensures that the adaptive sampling procedure operates on physically meaningful classifications.

The remainder of the paper is organised as follows. 
Section~\ref{sec:problem-statement} introduces the three flow configurations used to evaluate the new adaptive sampling method. 
Section~\ref{sec:Methodology} presents the full methodology, including the dataset construction, the classifier neural network, the KRnet-based generative model, and the adaptive sampling procedure. 
Section~\ref{sec:results} reports numerical results for each problem and demonstrates the effectiveness of the adaptive strategy in resolving nonlinear bifurcation boundaries with substantially fewer CFD simulations. 
Finally, Section~\ref{sec:conclusion} summarises the main findings and outlines potential directions for future work.

\section{Problem Statement}\label{sec:problem-statement}

The flow problems considered in this study are governed by the Navier--Stokes equations defined on a bounded spatial domain $D \subset \mathbb{R}^2$ and a temporal domain $\tau = (0, \infty)$. 
Given a time-dependent differential operator $\mathscr L$ and a parameter vector $\boldsymbol{\lambda} \in \mathbb{R}^n$, the governing system can be written as
\begin{equation}
	\mathscr L(\vec{u}(\mathbf{x}, t; \boldsymbol{\lambda})) = 0, \qquad \text{in } D \times \tau,
\end{equation}
\begin{equation}
	\vec{u}(\cdot, t; \boldsymbol{\lambda}) = g(t; \boldsymbol{\lambda}), \qquad \text{on } \partial D\times \tau,
\end{equation}
starting from a quiescent flow state at $t = 0$. 
The aim is to explore the bifurcation structure of the flow field $\vec{u}(\mathbf{x}, t; \boldsymbol{\lambda})$ as the parameters $\boldsymbol{\lambda}$ vary.

\subsection{Symmetry-breaking Channel Flow}\label{problem1}

The first problem is the classical symmetry-breaking channel flow with a symmetric expansion. 
The velocity $\vec{u} = [u_x, u_y]$ and pressure $p$ satisfy the Navier--Stokes equations in a rectangular channel with a $1{:}2$ expansion. 
The inlet, outflow, and wall boundary conditions are prescribed as
\begin{align}
	\vec{u}(-1,y) &= (1 - 4y^{2}, 0), 
	&& (-1,y),~|y| \le 0.5,  \\[3pt]
	\nu \frac{\partial u_x}{\partial x} &= p, 
	\frac{\partial u_y}{\partial x} = 0, && (16,y),~|y|\le 1,  \\[3pt]
	\vec{u} &= \vec{0}, 
	&& (x,\pm1),\ 0 \le x \le 16,\ (x,\pm0.5),\ -1 \le x \le 0, \\[2pt]
	&&& (0,y),\ 0.5 \le |y| \le 1. \notag
\end{align}

The base flow is symmetric with respect to the centerline $y = 0$, satisfying
\begin{equation}
	u_x(x, y) = u_x(x, -y), \qquad u_y(x, y) = -u_y(x, -y).
\end{equation}

The flow Reynolds number is defined as $\mathcal{R} = \frac{U L}{\nu}$, where $U$ is the maximum speed at the inlet, $L$ is the height of the channel at the outflow, and $\nu$ is the kinematic viscosity. 
As $\mathcal{R}$ increases beyond a critical value $\mathcal{R}^* \approx 220$, the symmetric steady solution becomes unstable and a pitchfork bifurcation occurs, giving rise to two stable asymmetric solutions that are mirror images of each other.

To simulate a controlled asymmetry, the inflow profile is perturbed as
\begin{equation}
	u_x(-1, y) = \delta \sin(2\pi y) + (1 - 4y^2),
\end{equation}
where $\delta$ denotes the imposed perturbation. 
The parameter vector for this problem is therefore becomes $\boldsymbol{\lambda} = [\mathcal{R},\, \delta].$

When $\delta = 0$, the flow remains symmetric until the critical Reynolds number is reached. 
For nonzero $\delta$, the perturbation biases the flow toward one of the asymmetric steady states. 
Representative symmetric and asymmetric steady-state flow fields for this configuration can be found in~\cite{silvester2026machine}. 

\subsection{Rayleigh--B\'enard Convection Problem}\label{problem2}

The first thermal convection problem corresponds to a classical Rayleigh--B\'enard configuration. Here, fluid motion arises due to buoyancy effects induced by a temperature difference between horizontal surfaces.  
A two-dimensional cavity of aspect ratio $1{:}2$ is considered, with insulated vertical boundaries at $x=0$ and $x=1$, and horizontal boundaries at $y=0$ and $y=2$. The lower boundary is heated while the upper boundary is cooled, creating a vertical temperature gradient capable of driving fluid motion.

When the imposed temperature difference is sufficiently small, thermal conduction dominates and the flow remains motionless. As the temperature increases, the conductive equilibrium eventually loses stability, giving rise to two possible steady convection states corresponding to clockwise and counterclockwise circulation. These two states form a pitchfork bifurcation pair, and representative flow fields illustrating each branch can be found in~\cite{silvester2026machine}.

The degree of instability of the system is characterized by the Rayleigh number, given as
\begin{equation}
	\mathrm{Ra} = \frac{g \, \beta (T_{\text{hot}} - T_{\text{cold}}) d^{3}}
	{\nu \, \epsilon},
\end{equation}
where $g$ is gravitational acceleration, $\beta$ is the thermal expansion coefficient, $d$ is the vertical distance between the horizontal walls, $\nu$ is the kinematic viscosity, and $\epsilon$ is the thermal diffusivity.  There is one more nondimensional parameter called the Prandtl number ($\mathrm{Pr}= \nu / \epsilon$). For water under the conditions considered here, the Prandtl number is $\mathrm{Pr} = 7.1$.

To model a physically realistic heating process, the temperature at the horizontal boundaries is ramped smoothly in time using a lifting function of the form $(1 - e^{-10t})$. An asymmetric perturbation $\delta$ is applied at the base of the cavity to examine how mild spatial variations in boundary temperature influence the pattern selection of the emerging convection. The resulting thermal boundary conditions are
\begin{align}
	T(\mathbf{x}, t) &= \left( \frac{1}{2} + \delta \sin(2\pi x) \right)
	(1 - e^{-10t}), 
	&& 0\leq x \leq 1,~y = 0,\ t > 0, \\
	T(\mathbf{x}, t) &= -\frac{1}{2}(1 - e^{-10t}), 
	&& 0\leq x \leq 1,~y = 2,\ t > 0,
\end{align}
with both vertical walls insulated. In summary, the bifurcation structure for this problem is investigated in the two-parameter space $(\mathrm{Ra}, \delta)$.

\subsection{A Differentially Heated Cavity Problem}\label{problem3}

This problem concerns natural convection in a differentially heated rectangular cavity, a configuration that undergoes a Hopf bifurcation from a steady flow to a time-periodic oscillatory regime at sufficiently large Rayleigh numbers. The physical motivation for this problem originates from the classical laboratory experiments of Elder~\cite{elder1965laminar}, who examined convection in a tall, narrow cavity filled with a highly viscous fluid. The working fluid in those experiments had an effective Prandtl number close to $\mathrm{Pr}\approx 1000$, characteristic of glycerol.

The computational setup considered here follows Elder’s geometry: a cavity of aspect ratio $0.051{:}1$, with the left vertical boundary ($x=0$) held at a higher temperature and the right boundary ($x=0.051$) cooled, while the horizontal walls remain thermally insulated. At low Rayleigh numbers the flow settles into a steady clockwise primary recirculation pattern. As the Rayleigh number increases past a critical value, the steady state loses stability through a Hopf bifurcation and the flow transitions to a time-periodic oscillatory motion. For $\mathrm{Pr}=1000$, this critical transition is observed numerically near $\mathrm{Ra}\approx 2.88\times 10^{9}$~\cite[p.~291]{Wubs2023}.
To investigate a broader class of fluids ranging from water to glycerol, the bifurcation structure is analyzed in the two-parameter space $(\mathrm{Ra},\mathrm{Pr})$ with $\mathrm{Pr}\in[7.1,1000]$. 

Although Elder’s experiments focused on a single highly viscous fluid with $\mathrm{Pr}\approx 1000$, the Rayleigh number at which the Hopf bifurcation occurs varies substantially across different fluids, and therefore depends on both $\mathrm{Ra}$ and $\mathrm{Pr}$. Extending these laboratory studies to fluids with moderate or low Prandtl numbers is difficult in practice, as the flow becomes highly sensitive to ambient disturbances. In contrast, numerical simulations permit systematic exploration of these regimes. By coupling high-fidelity computations with an adaptive, uncertainty-driven sampling strategy, the present work enables the mapping of stability boundaries across a wide range of $\mathrm{Pr}$, including parameter regimes that would be extremely difficult to reproduce experimentally.

\section{Methodology }\label{sec:Methodology}
We extend the machine-learning-based bifurcation classification strategy proposed in~\cite{silvester2026machine} into an \textit{adaptive} and \textit{data-efficient} approach. 
While the previous approach employed a neural network trained on a fixed dataset generated by carefully selecting parameter values in regions of the parameter space where transitions from stability to instability were expected, the present work begins with a uniformly sampled set of parameters covering the entire domain. 
An iterative procedure is then introduced that couples a classification network with a flow-based generative model to guide the sampling of new parameter points. 

\subsection{Generation of a Labeled Dataset}
\label{sec:dataset}

As in~\cite{silvester2026machine}, the learning scheme is based on a two-stage methodology. 
In the first stage, approximations of the flow solutions are computed by discretising the governing Navier--Stokes equations in space using mixed finite elements and in time using adaptive time stepping. In contrast to~\cite{silvester2026machine}, where parameter vectors $\boldsymbol{\lambda} = [\lambda_1, \lambda_2]$ were sampled in the neighbourhood of the expected bifurcation boundary, the present study begins with a set of uniformly distributed parameters covering the entire two-dimensional parameter domain. 
For each parameter combination, the corresponding solution is obtained starting from a quiescent base flow. 
A classification test is then applied to the computed flow field to determine whether it represents the base (symmetric) flow or whether the flow has undergone a bifurcation. 
The result is a labelled dataset 
\begin{equation}
\mathcal{D}_0 = \{ (\boldsymbol{\lambda}_i, \ell_i) \}_{i=1}^{n_0},
\end{equation}
where $\ell_i \in \{0,1\}$ denotes the label associated with the $i$th parameter combination, indicating nonbifurcated ($\ell_i=0$) or bifurcated ($\ell_i=1$) behaviour respectively. 
The labelling procedure follows exactly that described in~\cite{silvester2026machine} and is not repeated here. 
This initial dataset serves as the input to the classifier network described in Section~\ref{sec:classifier}, which is subsequently refined through adaptive sampling using the generative model introduced in Section~\ref{sec:KRnet}.

\subsection{Classifier Network for Bifurcation Prediction}
\label{sec:classifier}

The second stage of the methodology corresponds to the generation of a surrogate model for the bifurcation structure using a neural network classifier. 
The network architecture follow exactly those described in~\cite{silvester2026machine}. 

Before training, parameter samples obtained from the labeled dataset 
$\mathcal{D}_0 = \{ (\boldsymbol{\lambda}_i, \ell_i) \}_{i=1}^{n_0}$
are mapped from the physical parameter domain 
$[\boldsymbol{\lambda}_{\min}, \boldsymbol{\lambda}_{\max}]^2$
to a normalized domain $[-1,1]^2$ using an affine transformation,
\begin{equation}\label{eq:3.1}
	\boldsymbol{x}_i = 2\,\frac{\boldsymbol{\lambda}_i - \boldsymbol{\lambda}_{\min}}{\boldsymbol{\lambda}_{\max} - \boldsymbol{\lambda}_{\min}} - 1,
\end{equation}
where $\boldsymbol{\lambda}_{\min}$ and $\boldsymbol{\lambda}_{\max}$ denote the lower and upper bounds of the parameter space, respectively.  
Unlike the normalization strategy adopted in~\cite{silvester2026machine}, where each parameter component was standardized by subtracting its mean and dividing by its standard deviation, 
the present work employs a fixed affine scaling to the interval $[-1,1]^2$.
This approach ensures consistency with the domain required by the generative model and simplifies uniform numerical treatment of all parameters.
The normalized dataset is expressed as
\begin{equation}
\mathcal{D}_0^{(x)} = \{ (\boldsymbol{x}_i, \ell_i) \}_{i=1}^{n_0}.
\end{equation}

The input to the classifier is the two-dimensional normalized parameter vector
\[
\boldsymbol{x} = [x_1, x_2],
\]
and the output is a probability vector 
\(\mathbf{p} = [p_1, p_2]\),
representing the predicted likelihoods of the bifurcated and base (nonbifurcated) flow states, respectively.  
The network consists of a single hidden layer with 32 neurons and employs the standard sigmoid activation function.  
The hidden layer produces a two-component output vector $[r_1, r_2]$, which is converted into the probability vector $\mathbf{p}$ via a numerically stable softmax function:
\[
r_{*} = \max\{r_1,r_2\},\qquad p_j = \frac{e^{r_j-r_{*}}}{e^{r_1-r_{*}} + e^{r_2-r_{*}}}, \qquad j = 1,2,
\]
ensuring that $p_1 + p_2 = 1$.

For each labeled sample, the target output vector is defined as
\[
\boldsymbol{y}=[y_1, y_2] =
\begin{cases}
	[1, 0], & \text{if the flow is bifurcated,}\\[4pt]
	[0, 1], & \text{if the flow is nonbifurcated.}
\end{cases}
\]
The optimal classifier parameters ($\boldsymbol{\theta}_c^{*}$) are obtained by solving
\begin{equation}\label{eq:3:2}
	\boldsymbol{\theta}_c^{*} = \arg \min_{\theta_c}	\mathcal{L}_{\mathrm{C}}(\boldsymbol{\theta}_c),
\end{equation}
where $\boldsymbol{\theta}_c$ denotes the classifier parameters and $\mathcal{L}_{\mathrm{C}}(\boldsymbol{\theta}_c)$ denotes a cross-entropy loss function:
\begin{equation}\label{eq:3.3}
	\mathcal{L}_{\mathrm{C}}(\boldsymbol{\theta}_c) = - y_1 \ln(p_1) - y_2 \ln(p_2).
\end{equation}

In contrast to the neural network used for classification previously, the present study employs this network within an adaptive learning loop. 
The probabilistic output of the classifier is further utilized to quantify prediction uncertainty using the Shannon entropy,
\begin{equation}\label{eq:3.4}
H(\boldsymbol{x}) = - \sum_{j=1}^{2} p_j(\boldsymbol{x}) \log p_j(\boldsymbol{x}),
\end{equation}
which provides a scalar measure of the classifier’s confidence.  
Low entropy values ($H \approx 0$) correspond to confident predictions, while high values ($H \approx \log 2$) indicate ambiguity, typically near the bifurcation boundary.  
These regions of high entropy identify uncertain parts of the parameter space and are subsequently used to guide adaptive sampling through the generative model described in Section~\ref{sec:KRnet}.
While the classifier operates in the two-dimensional normalized parameter space, 
the subsequent generative model is formulated in a general $d$-dimensional setting 
to maintain scalability and notational generality.

\subsection{Adaptive Sampling with Flow-Based Deep Generative Model}
\label{sec:KRnet}

The second component of the proposed approach is a flow-based deep generative model, referred to as KRnet~\cite{tang2020deep,MR4385873}.
KRnet is a type of normalizing flow that learns an explicit probability density function (PDF) by constructing an invertible mapping between a simple latent space and a complex target space.
This invertibility enables both the evaluation of sample likelihoods and the efficient generation of new samples. Compared with adversarial or variational approaches~\cite{gulrajani2017improved,goodfellow2014,kingma2013auto}, normalizing flow models provide a tractable likelihood formulation, allowing their parameters to be optimized directly through likelihood maximization.
However, since the parameter space in the present framework is bounded, the probability distribution to be modeled is supported on a compact domain, whereas flow-based generative models are generally defined over the entire space.
To handle the bounded nature of the parameter domain, a smooth bijective mapping is introduced to transform it into an unbounded space where the flow transformation is applied.

Let $\boldsymbol{X} \in \mathbb{R}^d$ denote a random variable in the target parameter space with an unknown probability density function $p_{\boldsymbol{X}}(\boldsymbol{ x})$, and let $\boldsymbol{Z} \in \mathbb{R}^d$ represent a latent random variable drawn from a known prior distribution $p_{\boldsymbol{Z}}(\boldsymbol{z})$, typically a standard Gaussian. 
Flow-based generative models construct an invertible mapping
\begin{equation}\label{eq:3:5}
	\boldsymbol{z} = f(\boldsymbol{x}),
\end{equation}
which transforms samples from the target space to the latent space.
The inverse mapping $f^{-1}$ enables the generation of new samples in the target space by transforming latent samples drawn from the prior distribution back to the data space.

According to the change of variables, the probability density function of $\boldsymbol{X} = f^{-1}(\boldsymbol{Z})$ can be expressed as
\begin{equation}\label{eq:3:6}
	p_{\boldsymbol{X}}(\boldsymbol{x})
	= p_{\boldsymbol{Z}}\left(f(\boldsymbol{x})\right)
	\left|\det\nabla_{\boldsymbol{x}}f(\boldsymbol{x}) \right|.
\end{equation}

Here, the mapping $f$ is realized using the KRnet architecture, which employs a triangular transformation based on the Knothe--Rosenblatt rearrangement~\cite{MR2607321}. This design makes KRnet easily scalable to higher-dimensional sampling problems, which is particularly advantageous when the parameter space extends beyond two dimensions.

Let $f_{\boldsymbol{\theta}_g}(\cdot)$ denote the invertible transport map induced by KRnet, parameterized by $\boldsymbol{\theta}_g$. 
An explicit  probability density function based on this mapping can be written as
\begin{equation}\label{eq:3:7}
	p_{\boldsymbol{X}}(\boldsymbol{ x};\boldsymbol{\theta}_g)
	= p_{\boldsymbol{Z}}\left(f_{\boldsymbol{\theta}_g}(\boldsymbol{ x})\right)
	\left|\det\nabla_{\boldsymbol{x}} f_{\boldsymbol{\theta}_g}(\boldsymbol{x})\right|,
\end{equation}
where $p_{\boldsymbol{Z}}$ denotes the prior density in the latent space, typically chosen as a standard Gaussian distribution. 
Due to the invertibility of $f_{\boldsymbol{\theta}_g}$, samples from this model can be efficiently generated by sampling $\boldsymbol{Z}$ from $p_{\boldsymbol{Z}}$ and applying the inverse transformation,
\[
\boldsymbol{X} = f_{\boldsymbol{\theta}_g}^{-1}\left(\boldsymbol{Z}\right).
\]

Since, the normalized parameter space $\Omega =: [-1,1]^d$,  is bounded and the KRnet mapping $f_{\boldsymbol{\theta}_g}$ imposes no restriction on the range of the transformed variables; consequently, both $\boldsymbol{X}$ and $\boldsymbol{Z}$ are defined over $\mathbb{R}^d$.
Therefore, a bijective transformation is introduced to map the bounded normalized parameter domain to an unbounded space. Let us assume $S = (-(1+\varepsilon),\, (1+\varepsilon))^d$, with $0<\varepsilon< \infty$ such that $  \Omega \subset S$. For each dimension of $\boldsymbol{x} =[x_1,x_2,\cdots,x_d]$, a smooth logarithmic mapping is defined as

\begin{equation}\label{eq:3:8}
	y = \ell(x)
	= \log \left(\frac{x + (1+\varepsilon)}{(1+\varepsilon) - x}\right),
	\qquad 
	x = \ell^{-1}(y)
	= (1+\varepsilon)\frac{e^{y} - 1}{e^{y} + 1},
\end{equation}
where $\varepsilon>0$ defines a small buffer that slightly extends the domain to ensure numerical stability. 
This transformation establishes a one to one correspondence between $x \in (-(1+\varepsilon),\, (1+\varepsilon))$ and $y \in (-\infty,\, \infty)$, enabling the application of flow transformations defined on the whole space.

Let $\boldsymbol{\ell}(\boldsymbol{ x}) : S \rightarrow \mathbb{R}^d$ denotes the $d$-dimensional extension of this mapping, applied componentwise as
\begin{equation}\label{eq:3.9}
\boldsymbol{\ell}(\boldsymbol{ x}) =
\left[\ell(x_1),\, \ell(x_2),\, \ldots,\, \ell(x_d)\right]^{\!\top}.
\end{equation}
The overall invertible transformation used in KRnet can then be expressed as
\begin{equation}\label{eq:3.10}
	\boldsymbol{z} 
	= f_{\boldsymbol{\theta}_g}\!\big(\boldsymbol{\ell}(\boldsymbol{x})\big),
\end{equation}
which defines the composite mapping from the bounded parameter space to the latent space. 
The corresponding probability density function in the parameter domain is given by
\begin{equation}\label{eq:3.11}
p_{\boldsymbol{X}}^{(\boldsymbol{\ell})}(\boldsymbol{x}; \boldsymbol{\theta}_g)
	= p_{\boldsymbol{X}}(\boldsymbol{\ell}(\boldsymbol{x});\boldsymbol{\theta}_g)
	\left|\det\nabla_{\boldsymbol{x}}\boldsymbol{\ell}(\boldsymbol{x})\right|,
\end{equation}
where the last Jacobian term accounts for the logarithmic mapping introduced in~\eqref{eq:3.9} and the support of $p_{\boldsymbol{X}}^{(\boldsymbol{\ell})}(\boldsymbol{x}; \boldsymbol{\theta}_g)$ is $S$.
This formulation allows KRnet to approximate probability densities defined on compact parameter domains while retaining the mathematical tractability and sampling efficiency of normalizing flows on $\mathbb{R}^d$.

In the proposed adaptive sampling approach, KRnet serves as a data-driven generator that learns to produce new parameter samples in regions where the classifier exhibits higher uncertainty. 
Let $\boldsymbol{\theta}_c$ denote the parameters of the classification network, and let $\mathbf{p}(\boldsymbol{x};\boldsymbol{\theta}_c) = [p_1, p_2]$ denote the softmax probabilities corresponding to the two flow states. 
The uncertainty at each parameter point is quantified using the Shannon entropy:
\begin{equation}\label{eq:3.12}
	w(\boldsymbol{x})
	= -p_1\log(p_1) - p_2\log(p_2),
\end{equation}
where higher entropy values indicate regions close to the bifurcation boundary. 
KRnet is trained to adaptively focus on these regions by minimizing a weighted negative log-likelihood (WNLL) loss function:
\begin{equation}\label{eq:3.13}
\mathcal{L}_{w}(\boldsymbol{\theta}_g)
	= -\frac{1}{N}\sum_{i=1}^{N}
	w(\boldsymbol{x}_i)
	\log p_{\boldsymbol{X}}^{(\boldsymbol{\ell})}(\boldsymbol{x}_i; \boldsymbol{\theta}_g),
\end{equation}
where $p_{\boldsymbol{X}}^{(\boldsymbol{\ell})}$ is given by~\eqref{eq:3.11}. The optimal parameters of the generative model are obtained by solving
\begin{equation}\label{eq:3.14}
	\boldsymbol{\theta}_g^{*} 
	= \arg \min_{\boldsymbol{\theta}_g} \, \mathcal{L}_{w}(\boldsymbol{\theta}_g),
\end{equation}
where $\mathcal{L}_{w}$ is the weighted negative log-likelihood loss function defined in~\eqref{eq:3.13}.

The samples $\boldsymbol{x}_i$ are drawn uniformly from the normalized parameter space $[-1,1]^d$, and their weights $w(\boldsymbol{x}_i)$ are computed using the current classifier predictions. 
By emphasizing high-uncertainty points through entropy weighting, KRnet learns a sampling distribution that evolves adaptively with the classifier’s decision boundary. 
The choice of the weighted negative log-likelihood loss function ($\mathcal{L}_{w}$) for training KRnet is motivated by its ability to highlight regions of high classification uncertainty, which are most informative for identifying the bifurcation boundary. 
By minimizing the negative weighted log-likelihood, we effectively maximize the weighted log-likelihood, where the weights defined in~\eqref{eq:3.12} correspond to the Shannon entropy of the classifier output. 
These weights attain their maximum values when the predicted probabilities are nearly equal (\(p_{1}, p_{2} \approx 0.5\)), corresponding to points of maximum uncertainty near the bifurcation boundary. 
Consequently, larger weights assign greater importance to these ambiguous samples during training, compelling KRnet to allocate higher probability density to such regions in order to minimize the loss. 
This optimization mechanism naturally encourages the generative model to place more probability mass in regions where the classifier shows the highest uncertainty. Consequently, most of the samples drawn from the trained KRnet lie near the bifurcation boundary, while far fewer points are produced in areas that the classifier already labels with confidence. In this way, the method yields an efficient and adaptive sampling strategy for detecting boundaries.

Once trained under this objective, KRnet provides an adaptive sampling distribution that inherently focuses on high-uncertainty regions. New samples are generated by drawing $\boldsymbol{z}$ from the prior distribution $p_{\boldsymbol{Z}}$ and mapping them to the parameter domain using the inverse transformation of the mapping defined in~\eqref{eq:3.10}. The resulting samples are expected to lie within the physical bounds of the problem domain. In practice, we observed that nearly all generated samples remained within the admissible range, with at most one or two points falling slightly outside for a single test case. Therefore, no explicit filtering step was implemented in the present work. The generated samples are subsequently evaluated using CFD simulations to obtain their bifurcation labels. These labeled data points are then used to update the training dataset and retrain the classifier in the next adaptive iteration. The dataset is progressively expanded, as described in Section~\ref{sec:adaptive_sampling}. Through this iterative exchange between the classifier and KRnet, the sampling distribution gradually concentrates around the true bifurcation boundary, improving data efficiency and reducing the number of high-fidelity simulations required.

\subsection{Adaptive Sampling Procedure}
\label{sec:adaptive_sampling}

We now describe the adaptive sampling strategy employed to iteratively refine the training dataset in the parameter space. The key objective of this procedure is to improve the sampling efficiency by directing new samples toward regions of higher uncertainty, as identified by the classifier.
We use the following adaptive update strategy:

\vspace{0.2cm}
\noindent \textbf{Adaptive sampling with merging:} new samples are gradually added to the existing dataset, thereby expanding the training set in regions of high uncertainty. Let the initial labeled dataset be denoted by
\[
\mathcal{D}_0 = \{ (\boldsymbol{\lambda}_i^{(0)}, \ell_i^{(0)}) \}_{i=1}^{n_0},
\]
where $\boldsymbol{\lambda}_i^{(0)}$ are parameter samples uniformly distributed in the \textit{physical} parameter space defined by the problem bounds, and $\ell_i^{(0)}$ are the corresponding labels obtained from numerical simulations of the governing equations. 
Each physical parameter vector $\boldsymbol{\lambda}_i^{(0)}$ is subsequently mapped to a normalized coordinate 
$\boldsymbol{x}_i^{(0)}\in \Omega$ 
through an affine transformation defined in~\eqref{eq:3.1}. 
The normalized dataset 
\[
\mathcal{D}_0^{(x)} = \{ (\boldsymbol{x}_i^{(0)}, \ell_i^{(0)}) \}_{i=1}^{n_0}
\]
is then used to train the classifier network parameterized by $\boldsymbol{\theta}_c^{(0)}$ by minimizing the categorical cross-entropy loss defined in~\eqref{eq:3.3}. Using the trained classifier, the entropy values $w(\boldsymbol{x})$ are evaluated for a dense set of candidate points, and KRnet is trained with parameters $\boldsymbol{\theta}_g^{(0)}$ by minimizing the weighted negative log-likelihood loss defined in~\eqref{eq:3.13}. This establishes the initial adaptive sampling distribution.

Once the initial adaptive distribution is obtained at iteration \(k=0\), the procedure naturally extends to subsequent adaptive stages. 
At each iteration \(k+1\), the previously trained classifier and KRnet jointly define an updated sampling mechanism that further refines the exploration of the parameter space. 
The following steps describe the general procedure applied at an arbitrary iteration \(k\).

A set of normalized samples $\{ \boldsymbol{x}_{i}^{(k+1)} \}_{i=1}^{n_{\text{new}}}$ is obtained by $\ell^{-1}o f_{\boldsymbol{\theta}_g}^{-1}(\boldsymbol{z}^{(k+1)}_{i}; \theta_g^{*,(k)}),$ where
$\boldsymbol{z}^{(k+1)}_{i} $ are samples from the prior density, $p_{\boldsymbol{Z}}$ in the latent space.
These generated samples $\boldsymbol{x}^{(k+1)}_{i} \in [-1,1]^d$ are subsequently mapped back to the physical parameter space through an affine transformation,
\begin{equation}\label{eq:3.15}
\boldsymbol{\lambda}^{(k+1)}_{i}
= \boldsymbol{\lambda}_{\min} 
+ \frac{( \boldsymbol{x}^{(k+1)}_{i}+ 1)}{2} \, (\boldsymbol{\lambda}_{\max} - \boldsymbol{\lambda}_{\min}),~~~~i=1,\cdots,n_{\text{new}},
\end{equation}
where $\boldsymbol{\lambda}_{\min}$ and $\boldsymbol{\lambda}_{\max}$ denote the lower and upper bounds of the physical parameter domain.
The resulting physical parameters $\{ \boldsymbol{\lambda}^{(k+1)}_{i}\}_{i=1}^{n_{\text{new}}}$ are then used to perform direct numerical simulations of the governing equations, from which the corresponding bifurcation labels $\{\ell^{(k+1)}_{i}\}_{i=1}^{n_{\text{new}}}$ are obtained.
This produces an updated labeled dataset
\[
\mathcal{D}_{k+1} = \{ (\boldsymbol{\lambda}^{(k+1)}_{i}, \ell^{(k+1)}_{i}) \}_{i=1}^{n_{\text{new}}}.
\]
The classifier is then retrained using 
\[
\mathcal{D}_{k+1} ^{(x)} = \mathcal{D}_{k} ^{(x)} \cup \{ (\boldsymbol{ x}^{(k+1)}_{i}, \ell^{(k+1)}_{i}) \}_{i=1}^{n_{\text{new}}}.
\]

After retraining, entropy values are updated, and KRnet is trained again to refine the sampling distribution.
This iterative procedure is repeated until a prescribed adaptivity level is reached or the uncertainty near the bifurcation boundary becomes negligible.

For completeness, we briefly summarize the initialization of both models, as this determines how their parameters evolve across adaptive stages. The classifier network is initialized only once at the beginning of the adaptive procedure: all hidden-layer weights and biases are drawn from a zero-mean normal distribution, while the output-layer weights and biases are drawn from a uniform distribution on $[-1,1]$. KRnet is likewise initialized only once. The neural networks appearing in the affine coupling layers use Glorot normal initialization~\cite{glorot2010understanding} for their weights, and zero initialization for their biases. Once this initialization is performed, the parameters obtained at each stage are retained and used as the starting point for the subsequent adaptive stage.
\begin{algorithm}[!t]
	\caption{Adaptive sampling with merging}
	\label{alg:AS-M-DAS}
	\begin{algorithmic}[1]
		\Require Initial labeled set $\mathcal{D}_0=\{(\boldsymbol{\lambda}^{(0)}_i,\ell^{(0)}_i)\}_{i=1}^{n_0}$;
		maximum adaptive stages $n_{\text{adap}}$; classifier epochs $n_e^{(c)}$; KRnet epochs $n_e^{(g)}$; KRnet batch size $m$; candidate set size $n_{\text{cand}}$; new samples $n_{\text{new}}$
		\State {// Normalize once}
		\State Form $\mathcal{D}^{(x)}_0$ from $\mathcal{D}_0$ using~\eqref{eq:3.1}
		\For{$k = 0 : n_{\text{adap}}-1$} 
		
		\Statex {// Train classifier on current set $\mathcal{D}^{(x)}_k$}
		\For{$i = 1 : n_e^{(c)}$} 
		\State Randomly permute $\mathcal{D}^{(x)}_k$
		\For{each $(\boldsymbol{x},\ell)$ in the permuted $\mathcal{D}^{(x)}_k$}
		\State Update classifier parameters by \emph{gradient descent} of $\mathcal{L}_{\mathrm{C}}(\boldsymbol{\theta}_c)$ \hfill (see~\eqref{eq:3.3})
		\EndFor
		\EndFor
		
		\Statex {// Compute entropy weights on a candidate set}
		\State Sample a candidate pool $\mathcal{S}_{\Omega,k}\subset \Omega$ of size $n_{\text{cand}}$ (e.g., uniform random)
		\State Evaluate classifier probabilities on $\mathcal{S}_{\Omega,k}$ and compute $w(\boldsymbol{x})$ via~\eqref{eq:3.12}
		
		\Statex {// Train KRnet}
		\For{$i = 1 : n_e^{(g)}$} 
		\For{$j $ steps} 
		\State Sample $m$ points from $\mathcal{S}_{\Omega,k}$
		\State Update KRnet parameters using the \emph{ADAM optimizer} to minimize $\mathbb{E}\big[\mathcal{L}_{w}(\boldsymbol{\theta}_g)\big]$ \hfill (see \eqref{eq:3.13})
		\EndFor
		\EndFor
		
		\Statex {// Generate and label $n_{\text{new}}$ new points }
		\State Generate $n_{\text{new}}$, $\boldsymbol{x}^{(k+1)}\in \Omega$ points through $p_{\boldsymbol{X}}^{(\boldsymbol{\ell})}(\boldsymbol{x}; \boldsymbol{\theta}_g)$ \hfill (see \eqref{eq:3.11})
		\State Map to physical parameters, $\boldsymbol{\lambda}^{(k+1)}$, using~\eqref{eq:3.15}, and obtain the label $\ell^{(k+1)}$
		
		\Statex {// Merging update}
		\State Set $\mathcal{D}^{(x)}_{k+1} \gets \mathcal{D}^{(x)}_k \cup \{(\boldsymbol{x}^{(k+1)},\ell^{(k+1)})\}$
		\EndFor
		\State {Output:} final classifier $\boldsymbol{\theta}_c^{*}$ and KRnet $\boldsymbol{\theta}_g^{*}$
	\end{algorithmic}
\end{algorithm}

\section{Results and Discussion}\label{sec:results}

In this section, we present a series of numerical experiments to show the effectiveness of the proposed adaptive sampling approach to classify the bifurcation boundary. We consider here three flow problems, spanning both symmetry-breaking pitchfork bifurcations and a Hopf bifurcation, thereby allowing the method to be analysed across different stability transitions. These include a symmetry-breaking channel flow, a Rayleigh--B\'enard convection problem, and a differentially heated cavity problem thoroughly discussed in Section~\ref{sec:problem-statement}. The classifier neural network employs a sigmoid-activation function with a softmax output, while KRnet uses the rectified linear unit (ReLU) activation function.

\subsection{Symmetry-Breaking Channel Flow}\label{problem1:discussion}

We begin with the symmetry-breaking channel-flow problem introduced in Section~\ref{problem1}, which is a useful test case for assessing the ability of the proposed adaptive sampling strategy. The transition from a symmetric base flow to asymmetric steady states is governed by the Reynolds number $(\mathcal{R})$ and the amplitude of an imposed inlet perturbation $(\delta)$. For this example, we train the classifier and KRnet using the procedure described in Section~\ref{sec:adaptive_sampling}, and compare the predicted bifurcation boundary obtained at different steps to highlight the progressive refinement achieved by the adaptive strategy.

\begin{figure}
	\centering	\includegraphics[width=0.95\linewidth]{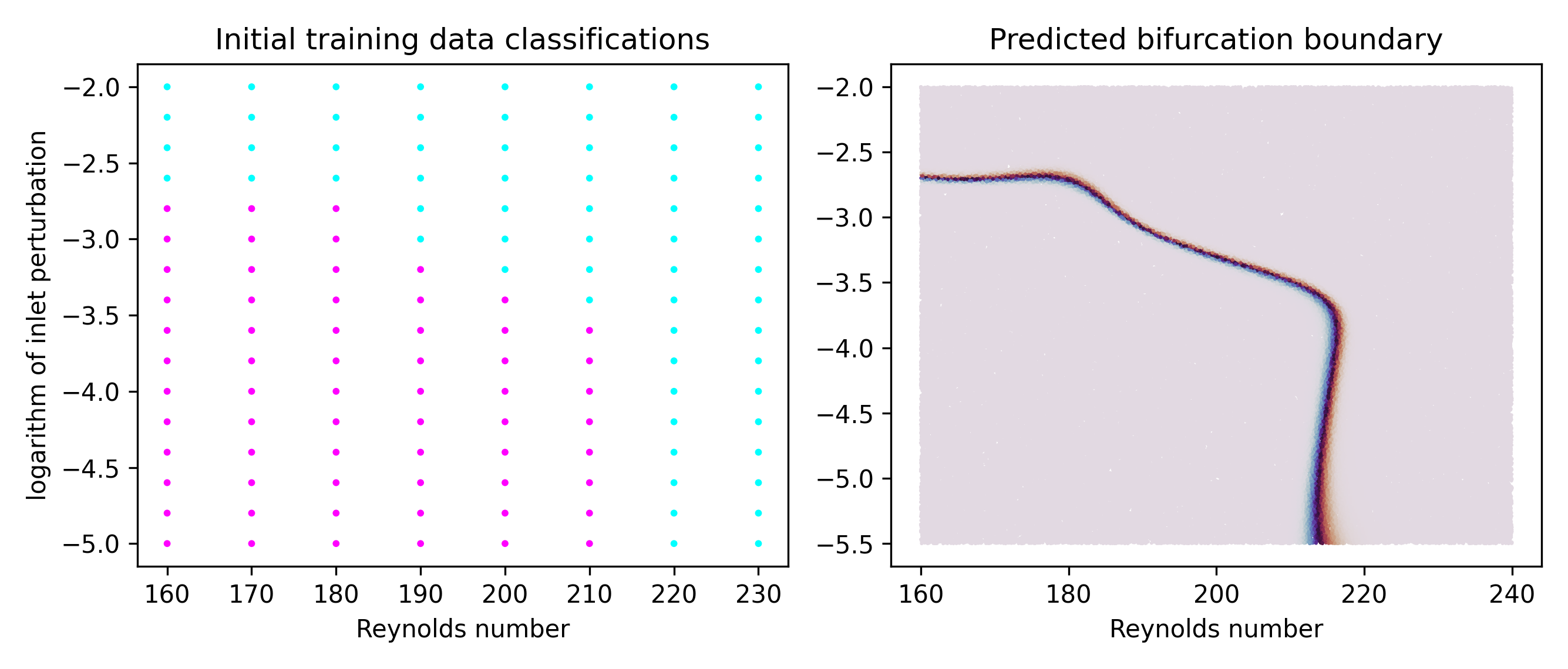}
	\includegraphics[width=0.95\linewidth]{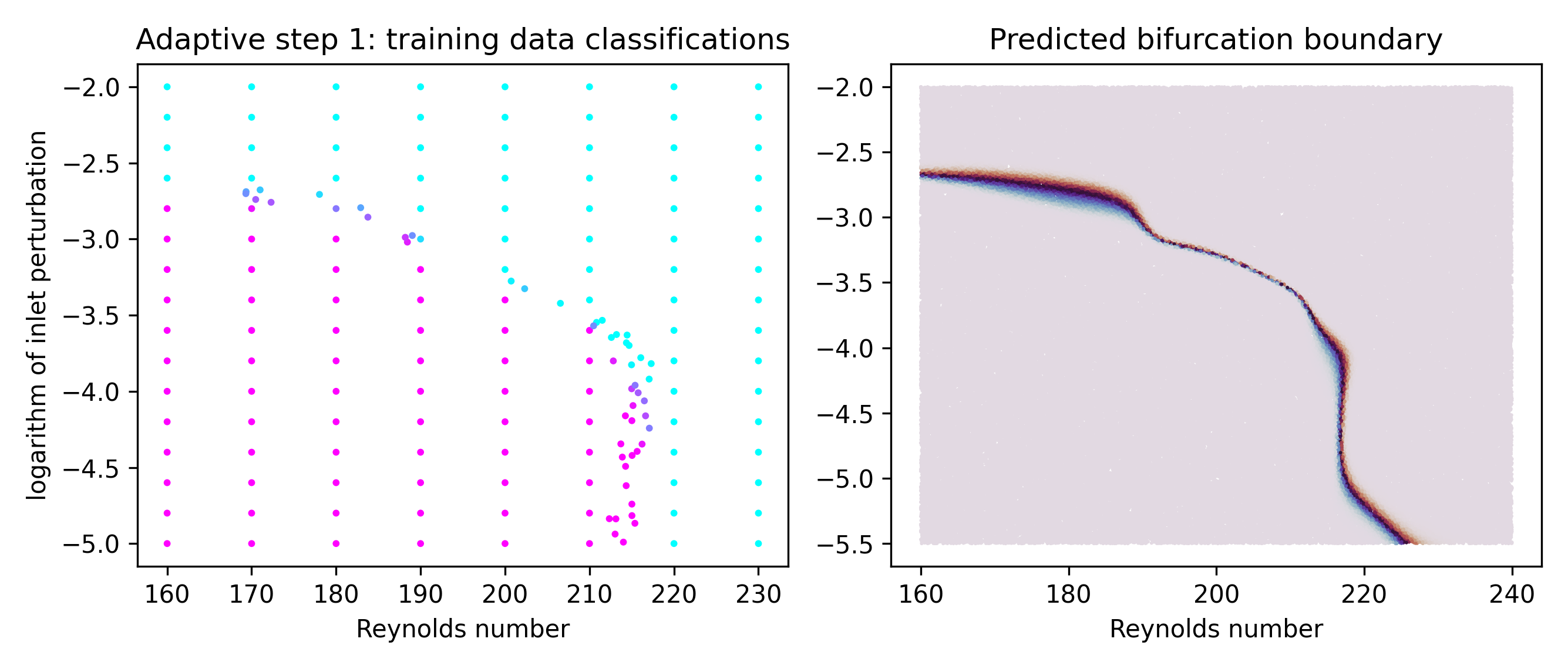}	\includegraphics[width=0.95\linewidth]{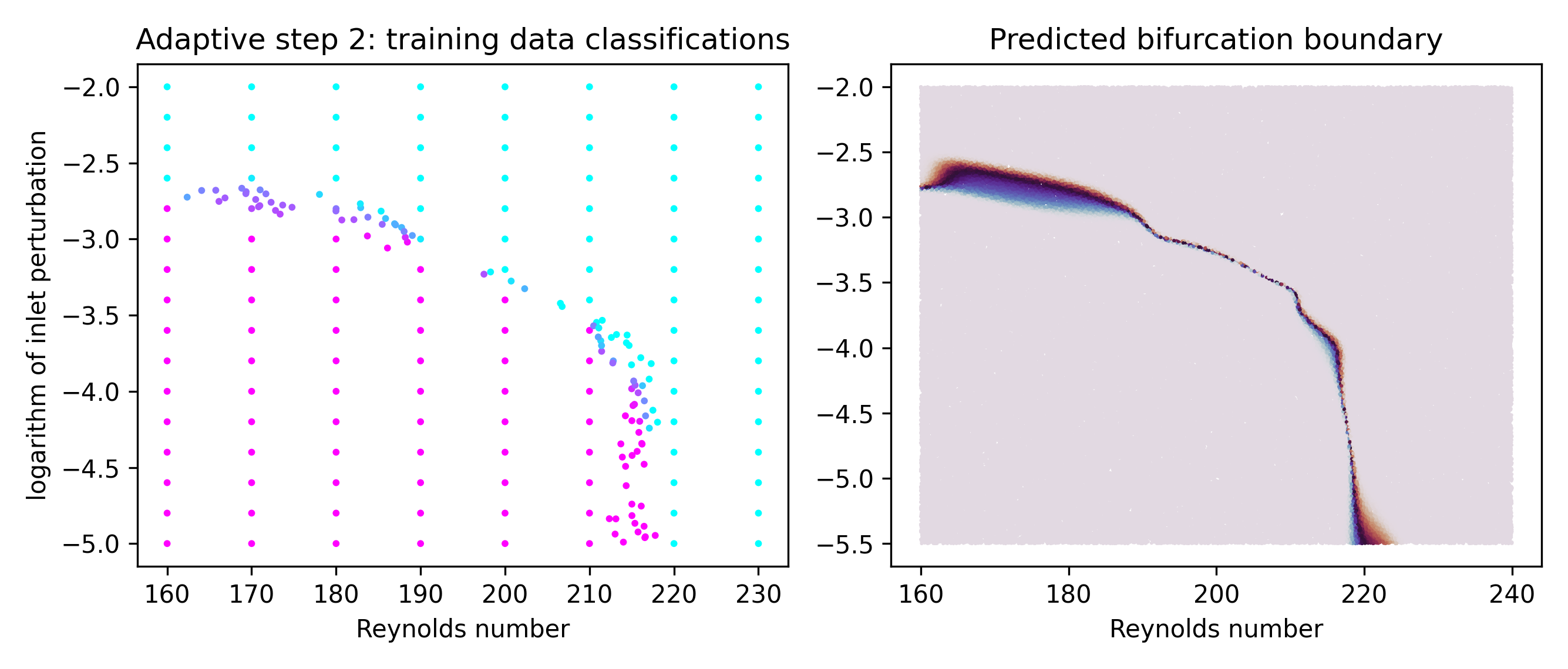}
	\caption{{Adaptive refinement of the training set and the resulting bifurcation boundary for the channel-flow problem. Each row shows the training data (left) and the classifier-predicted boundary (right) at the initial, first, and second refinement steps.}}
	\label{problem1_boundary_plots}
\end{figure}

For this problem, the initial dataset is generated on a uniform grid in the
$(\mathcal{R},\delta)$ parameter space, consisting of $8$ uniformly spaced
Reynolds numbers in the range $\mathcal{R}\in[160,230]$ and $16$ logarithmically
spaced inlet perturbations $\delta\in[10^{-5},10^{-2}]$, providing
$128$ labeled samples. These labels are obtained from Navier--Stokes
solutions computed using the IFISS software and classified according to the
symmetry criterion of~\cite{silvester2026machine}. The classifier neural network used in this example has a single hidden layer with $32$ neurons and a softmax
output layer, trained for $n_e^{(c)}=3000$ epochs using stochastic gradient
descent (batch size $1$, learning rate $0.15$). To construct the initial
training set for KRnet, a set of $n_{\text{cand}}=6000$ additional
parameter points sampled uniformly in the normalized domain is evaluated by the
classifier to compute the entropy weights $w(\boldsymbol{x})$. KRnet employs
$6$ affine coupling layers, each equipped with two fully connected subnetworks
of $24$ neurons per layer, and is trained using the WNLL loss with the ADAM
optimizer (batch size $500$, learning rate $10^{-3}$) for $n_e^{(g)}=1000$ epochs. In total, two adaptive refinement steps are carried out, and at each step the KRnet contributes $n_{\text{new}}=50$ additional parameter points to the training set.

Figure~\ref{problem1_boundary_plots} demonstrates how the adaptive procedure progressively refines the training set and show the associated bifurcation boundary for this problem. In each row, the left panel shows the classification of the training data in the $(\mathcal{R},\log_{10}\delta)$ space, where magenta points correspond to nonbifurcated solutions and cyan points correspond to bifurcated solutions. The top-left plot corresponds to the initial dataset, which forms a uniform grid and provides only a coarse sampling of the transition region. After the first and second adaptive adaptive refinements, additional samples generated by KRnet are incorporated and collect along the uncertain region between the two classes, leading to a much denser coverage of the bifurcation boundary. The right plots display the corresponding predicted bifurcation boundary, obtained by assessing the classifier on $2\times 10^{5}$ randomly sampled parameter points. As the adaptive steps proceed, the high-probability transition band becomes sharper and follows more closely the shape of the true bifurcation curve, demonstrating the ability of the classifier-KRnet approach to automatically focus sampling effort where it is most informative.

A closer inspection of the classification patterns reveals how the adaptive strategy responds to evolving regions of uncertainty. In the initial training dataset, the classifier exhibits its greatest ambiguity near $\mathcal{R}\approx 218$ and $\log_{10}(\delta)\in[-5,-3.5]$, where the bifurcation curve cannot be reliably captured by the coarse grid alone. Consequently, KRnet focuses new samples in this neighbourhood, generating points precisely along the transition curve and producing a sharper and more clear boundary after the first adaptive step. Next, we see the uncertainty towards larger perturbations, specifically for $\log_{10}(\delta)\in[-2.75,-2.5]$ and Reynolds numbers $\mathcal{R}\approx 170$--$190$. The second adaptive step again directs sampling to this newly uncertain region. These focused refinements collectively highlight the effectiveness of the adaptive sampling approach.

\begin{figure}[!t]
	\centering
	\includegraphics[width=\linewidth]{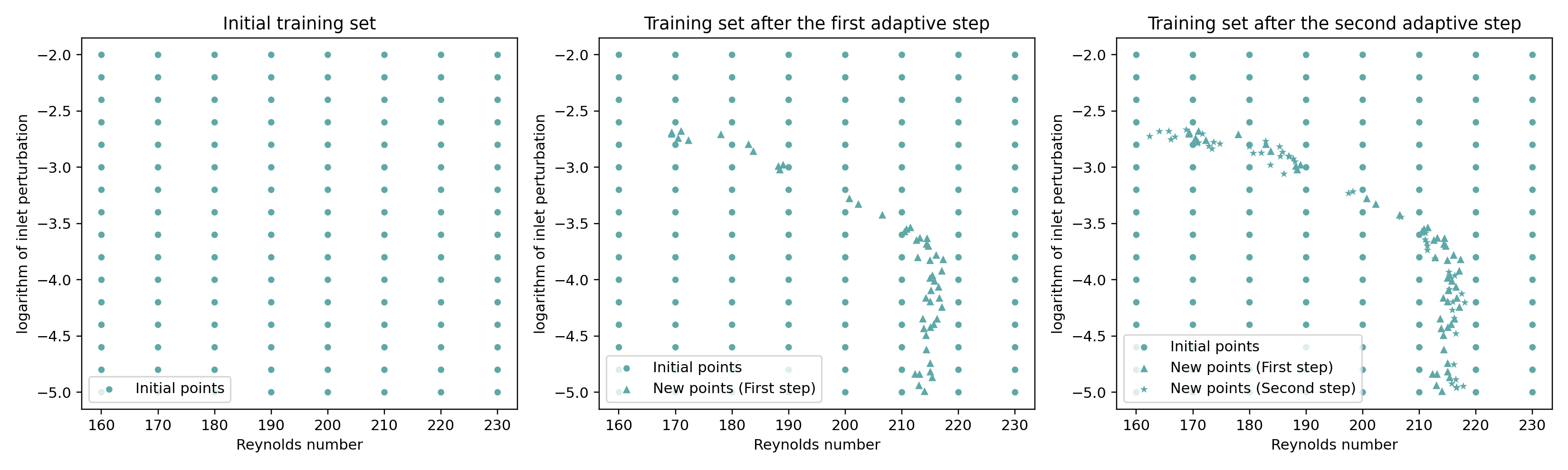}
	\caption{Training-set growth across two adaptive refinement steps for the symmetry-breaking channel flow problem.} \label{problem1_KRnet_generated_points}
\end{figure}

Figure~\ref{problem1_KRnet_generated_points} shows the KRnet-generated points at the first two adaptive refinement steps. As discussed thoroughly in the methodology section, KRnet naturally concentrates samples near regions of high classifier uncertainty, which here correspond to the bifurcation boundary. The left plot shows the initial training set consisting of 128 uniformly spaced samples in the $(\mathcal{R},\log_{10}(\delta))$ parameter domain. The middle plot shows the first adaptive update, where KRnet contributes 50 new points, all lying strictly inside the admissible domain. The right plot shows the second adaptive update, containing another 50 KRnet-generated points, all of which remain strictly within the domain. Since all points consistently lie within the admissible range, no projection or filtering step was required.

\begin{figure}[!t]
	\centering
	\begin{subfigure}{.5\textwidth}
		\centering
		\includegraphics[width=1\linewidth]{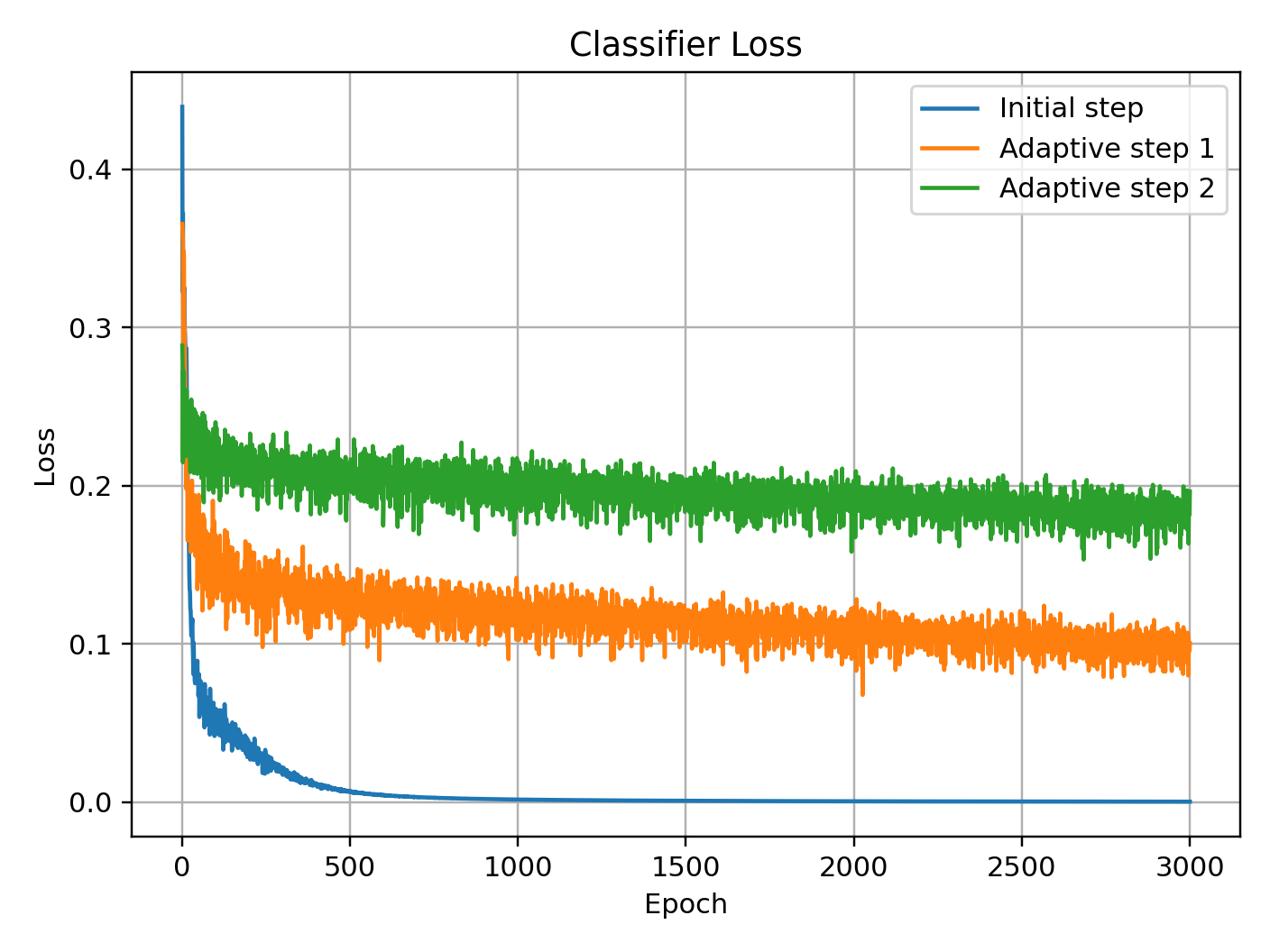}
	\end{subfigure}%
	\begin{subfigure}{.5\textwidth}
		\centering
		\includegraphics[width=1\linewidth]{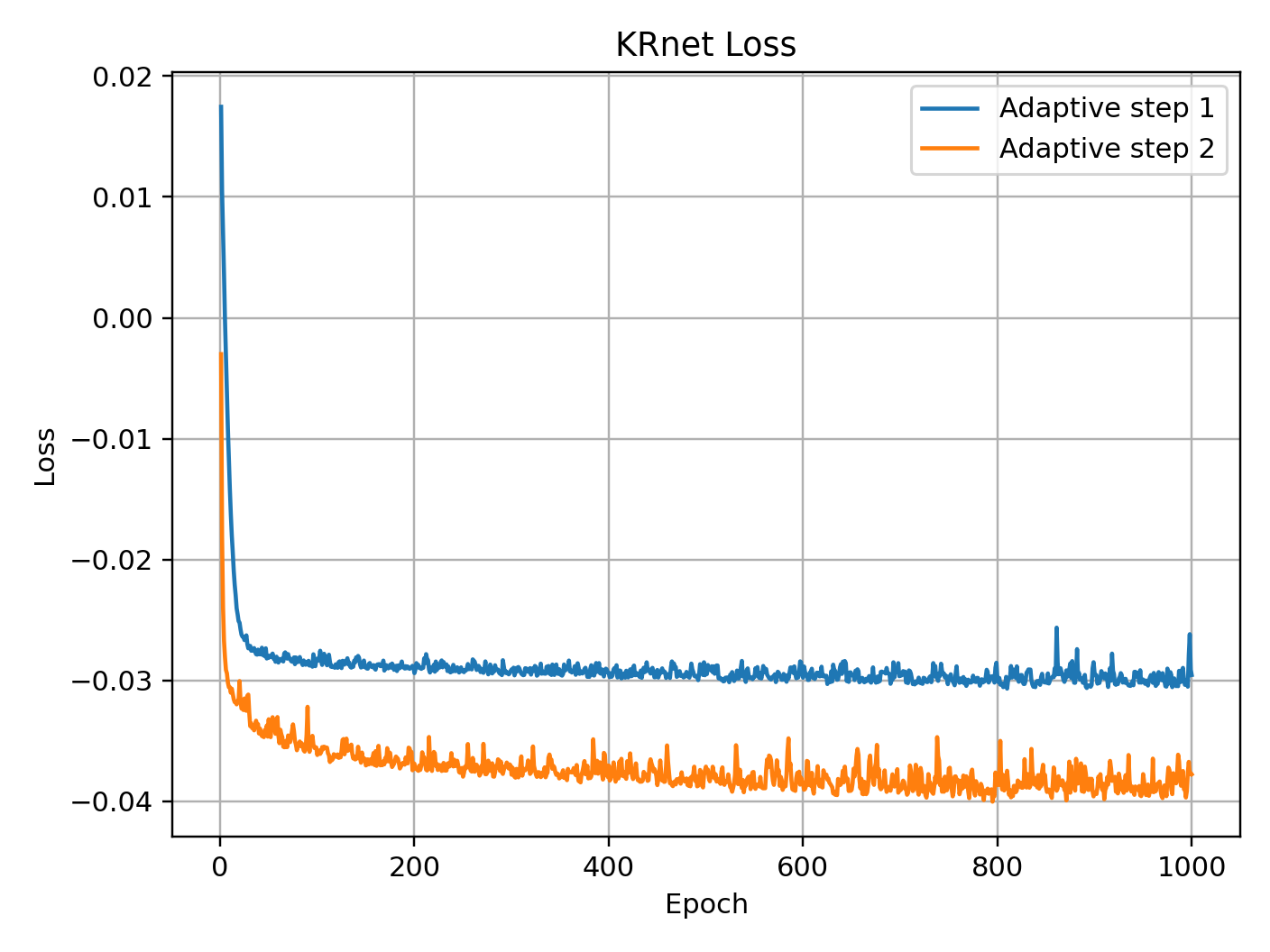}
	\end{subfigure}
	\caption{Comparison of loss evolution for the two learning components. Left: Classifier loss across the initial step and two adaptive refinement steps. Right: KRnet WNLL loss for the first two adaptive steps.}\label{problem1_losses}
\end{figure}

The plots in Figure~\ref{problem1_losses} present the classifier loss and KRnet WNLL loss across successive adaptive steps of the training process. For the classifier (left plot), all curves exhibit a sharp decrease in loss during the initial epochs, followed by a stable plateau where the loss remains nearly unchanged, indicating that the classifier completes most of its learning early and additional epochs do not lead to further improvements. The primary effect of adaptive refinement is a lower initial loss at the beginning of each new step, reflecting that the refined samples provide a better starting point for training, while the overall convergence behavior remains similar across steps. A comparable trend is observed in the KRnet loss (right plot), where both adaptive steps converge rapidly to nearly identical plateau values and show no significant differences in their long-term behavior. Together, these results indicate that adaptive refinement mainly enhances the initial training conditions for both components, whereas the eventual convergence pattern remains stable across adaptive steps.

 \subsection {Rayleigh--B\'enard convection problem}\label{problem2:discussion}
Next, we look at the Rayleigh--B\'enard convection problem described in Section~\ref{problem2}. This test case serves as a thermally driven analogue to the channel-flow example. Here, the transition from a motionless conductive state to a steady convective circulation is controlled by the Rayleigh number $(\mathrm{Ra})$ together with a small asymmetric thermal perturbation imposed along the heated lower surface. This perturbation biases the system toward one of the two symmetry-related convection states and therefore plays an essential role in shaping the bifurcation curve.

For this problem, the initial training set is generated by sampling the $(\mathrm{Ra},\delta)$ parameter domain on a uniform grid. Sixteen Rayleigh numbers are selected uniformly over the interval $\mathrm{Ra}\in[1300,1600]$, and eight perturbation amplitudes are constructed by choosing exponents uniformly in $[-16,-2]$ and taking the corresponding powers of ten, giving $\delta \in [10^{-16},10^{-2}]$. This results in a total of $128$ parameter pairs. For each pair, the governing equations are solved using the IFISS software, and the resulting solution is labeled as bifurcated or nonbifurcated using the stability criterion described in~\cite{silvester2026machine}. The classifier retains the same architecture as in the channel-flow case and is trained for $n_e^{(c)} = 3000$ epochs using stochastic gradient descent with batch size~$1$ and learning rate~$0.15$.
To train the KRnet model, a set of $n_{\text{cand}}=6000$ uniformly sampled parameter points in the normalized domain is evaluated by the classifier to compute entropy-based weights. KRnet again uses $6$ affine coupling layers, and for this problem it is trained for $n_e^{(g)} = 2000$ epochs using the WNLL loss fuction and the ADAM optimizer with batch size $500$ and learning rate $10^{-3}$. At each adaptive refinement step, KRnet contributes $n_{\text{new}} = 60$ new parameter points to the training set, and two such refinements are performed for this problem.

 \begin{figure}
	\centering
	\includegraphics[width=0.95\linewidth]{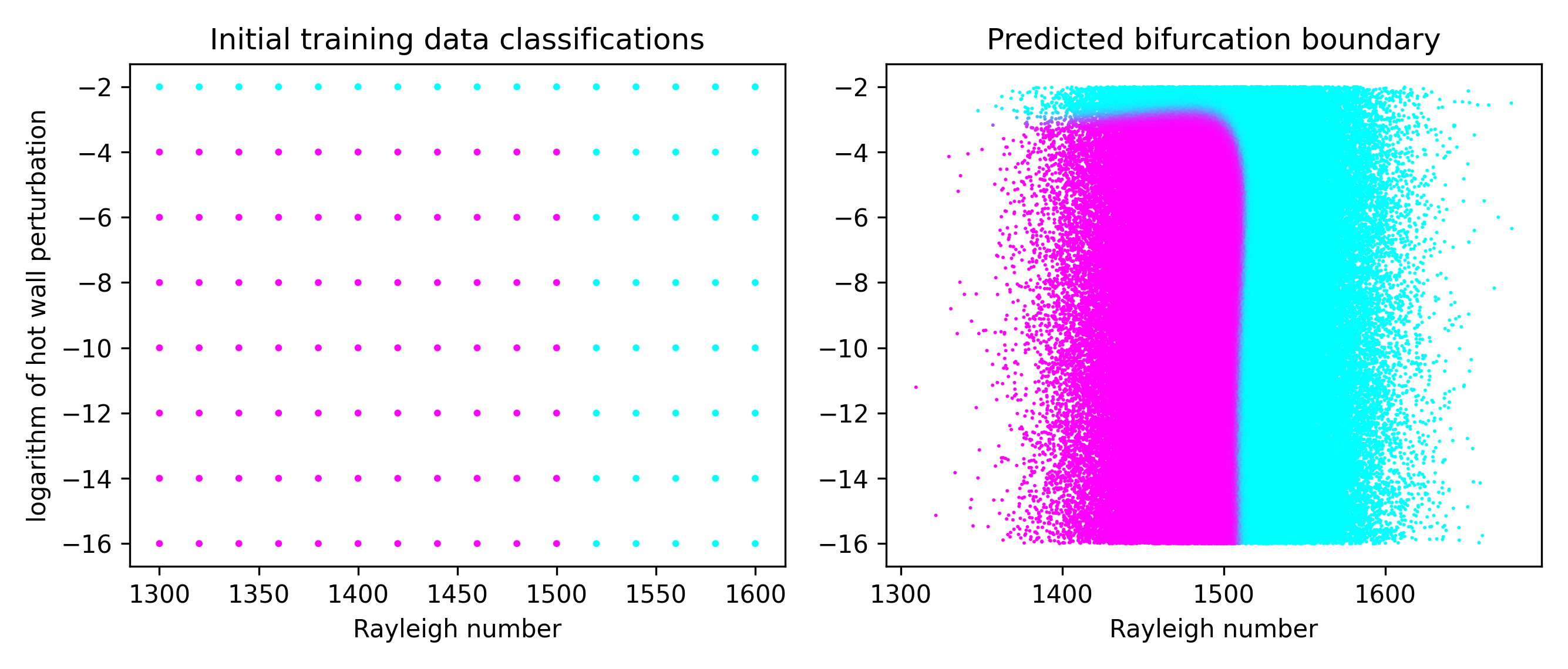}
	\includegraphics[width=0.95\linewidth]{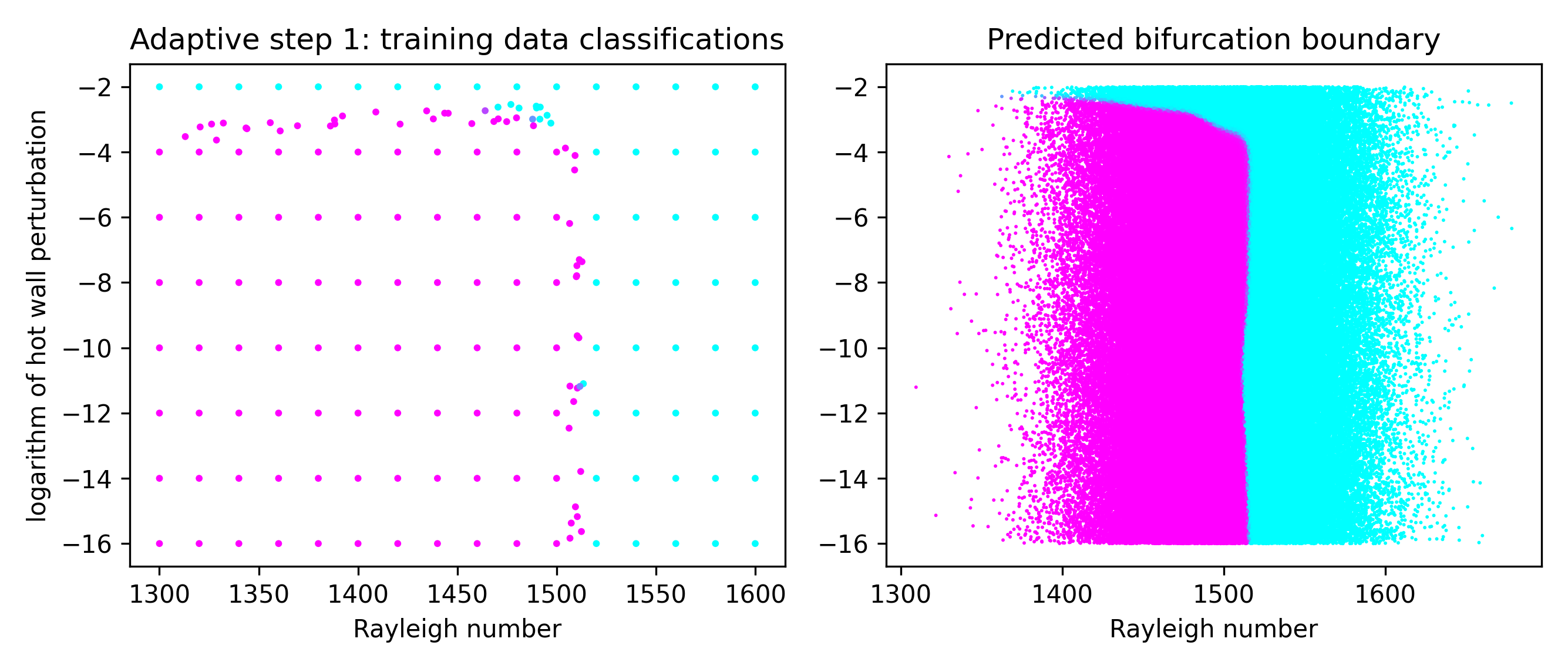}
	\includegraphics[width=0.95\linewidth]{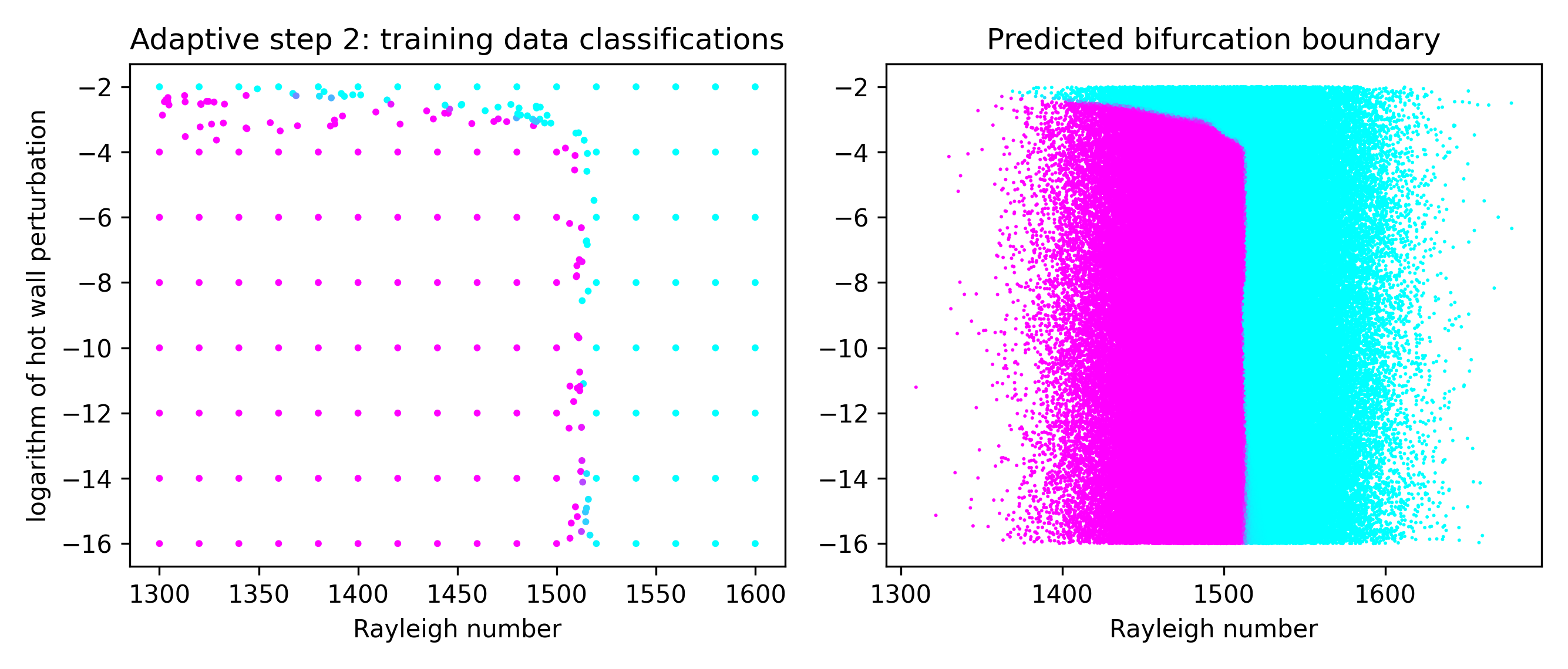}
	\caption{{Adaptive refinement of the training set and the resulting bifurcation boundary for the Rayleigh--B\'enard convection problem. Each row shows the training data (left) and the classifier-predicted boundary (right) at the initial, first, and second refinement steps.}}\label{problem2_boundary_plots}
\end{figure}

Figure~\ref{problem2_boundary_plots} shows how the adaptive sampling procedure refines the training set and inferred bifurcation boundary. In each row, the left plot shows the classified training samples in the $(\mathrm{Ra},\log_{10}\delta)$ space, with magenta points indicating nonbifurcated states and cyan points indicating bifurcated states. In the initial dataset, the two classes are not cleanly separated, and a blue band appears between the magenta and cyan regions, reflecting the classifier's high uncertainty in this poorly sampled portion of the parameter space. After each refinement step, the KRnet generates new samples that concentrate along this uncertain part, producing a progressively denser and more informative distribution of training points. The right plots display the corresponding classifier-predicted bifurcation boundary, obtained by evaluating the trained network on $2\times10^{5}$ randomly sampled parameter points.

A qualitative interpretation of the learned transition curve reveals several notable trends. For very small perturbations (large negative $\log_{10}\delta$), the boundary is nearly vertical, indicating that the onset of convection is governed primarily by the Rayleigh number when the bottom-wall heating is almost perfectly uniform. As $\delta$ increases beyond roughly $10^{-3}$, the transition curve begins to shift toward lower Rayleigh numbers, showing that even modest asymmetry in the imposed temperature profile can induce convection at earlier parameter values. Near the upper end of the sampled range ($\delta \approx 10^{-2}$), this shift becomes more pronounced, with the predicted critical Rayleigh number falling to values close to $1300$. Across all refinement steps, the adaptive procedure continues to position new samples smoothly along this rapidly changing region, yielding a sharper and more clear estimate of the bifurcation boundary.

 \subsection {Differentially heated cavity}\label{problem3:discussion}
We now turn to the differentially heated cavity problem introduced in Section~\ref{problem3}, which provides an example of a flow undergoing a Hopf bifurcation. In this case, the loss of stability from a steady circulation to a time-periodic oscillatory state is governed by the Rayleigh number $(\mathrm{Ra})$ for different types of fluids characterised by the Prandtl number $(\mathrm{Pr})$. 

For this example, the initial training set is formed by sampling eight fixed Prandtl numbers, representing a range of fluids from air ($\mathrm{Pr}=0.71$) and water ($\mathrm{Pr}=7.1$) to more viscous liquids ($\mathrm{Pr}=140,195,250,500$) and finally glycerol ($\mathrm{Pr}\approx 1000$). For each $\mathrm{Pr}$, eleven Rayleigh numbers are chosen uniformly in the interval $\mathrm{Ra}\in[2.7\times10^{9},\,3.2\times10^{9}]$, providing a total of $88$ initial parameter samples. Each solution is computed with IFISS and labeled according to whether the system settles to a steady state or exhibits time-periodic oscillations as in~\cite{silvester2026machine}. The classifier and KRnet architectures, as well as all training hyperparameters are taken to be the same as in problem~\ref{problem2:discussion}. At each refinement step, KRnet contributes $n_{\text{new}}=50$ new parameter points, which are then added to the training set for the subsequent classifier update.

\begin{figure}[!t]
	\centering
	\includegraphics[width=0.95\linewidth]{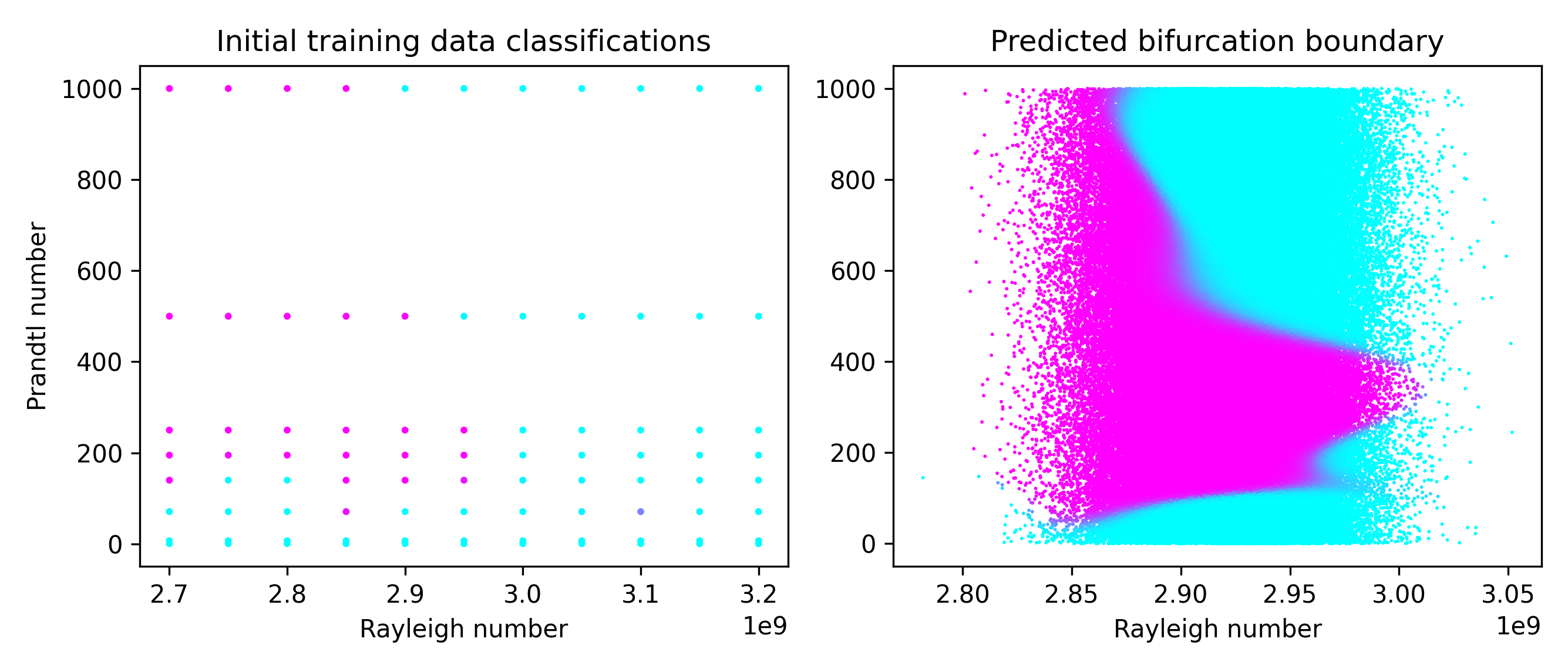}
	\includegraphics[width=0.95\linewidth]{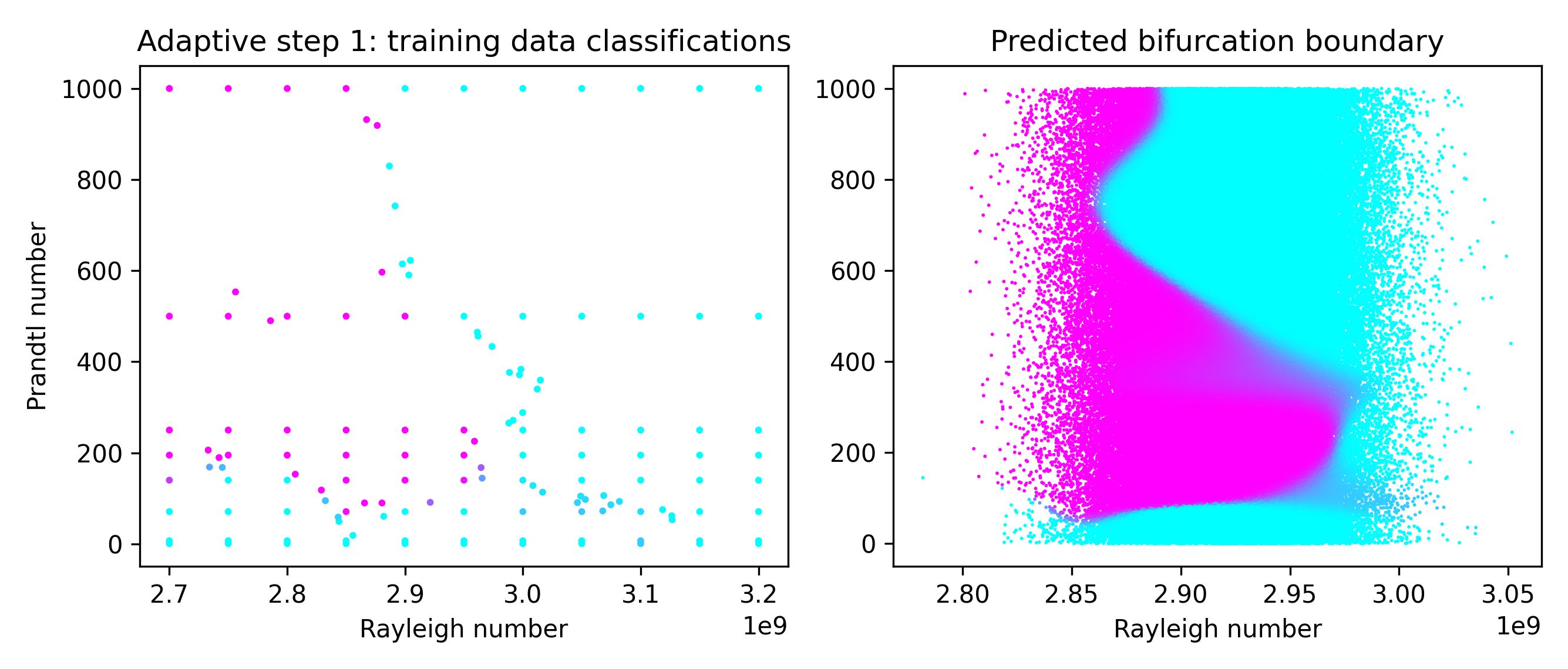}
	\caption{Adaptive refinement of the training set and the resulting bifurcation boundary for the differentially heated cavity problem. Each row shows the training data (left) and the classifier-predicted boundary (right) at the initial step and after the first adaptive refinement.}\label{problem3_boundary_plots_part1}
\end{figure}

Figure~\ref{problem3_boundary_plots_part1} shows the initial training data and the classifier-predicted bifurcation boundary for the differentially heated cavity problem. In the left plot, magenta points denote steady (nonbifurcated) solutions and cyan points denote periodic (bifurcated) ones. The initial grid provides only a coarse sampling of the $(\mathrm{Ra},\mathrm{Pr})$ domain, and the corresponding prediction in the right panel contains regions where the classifier is uncertain. One such instance is the point near $(\mathrm{Ra},\mathrm{Pr}) \approx (3.1\times10^{9},140)$, which appears as a blue dot in the probability field. This colour indicates that the classifier assigns a prediction close to $0.5$, indicating that, based on the initial training data, it cannot reliably determine whether this point corresponds to a steady or periodic state. Although this point was originally labelled as nonbifurcated by the IFISS simulation, its isolated position relative to the surrounding samples naturally leads to higher uncertainty in the learned decision boundary.
The effect of adaptive refinement is illustrated in the second row of Figure~\ref{problem3_boundary_plots_part1}. Here, KRnet generates additional parameter points in regions of high classifier uncertainty, which are then incorporated into the training set. After the first refinement step, several new samples are added in the region around the point that was previously ambiguous. With this additional local information, the classifier updates its decision and identifies that region as part of the bifurcated class, shown in the updated prediction by the cyan colour. This behaviour highlights one of the strengths of the adaptive strategy: the neural network interpolates the underlying CFD-generated labels and uses the refined sampling to resolve inconsistencies or isolated misclassifications that arise from the initial coarse grid.

\begin{figure}[!t]
	\centering
	\includegraphics[width=0.95\linewidth]{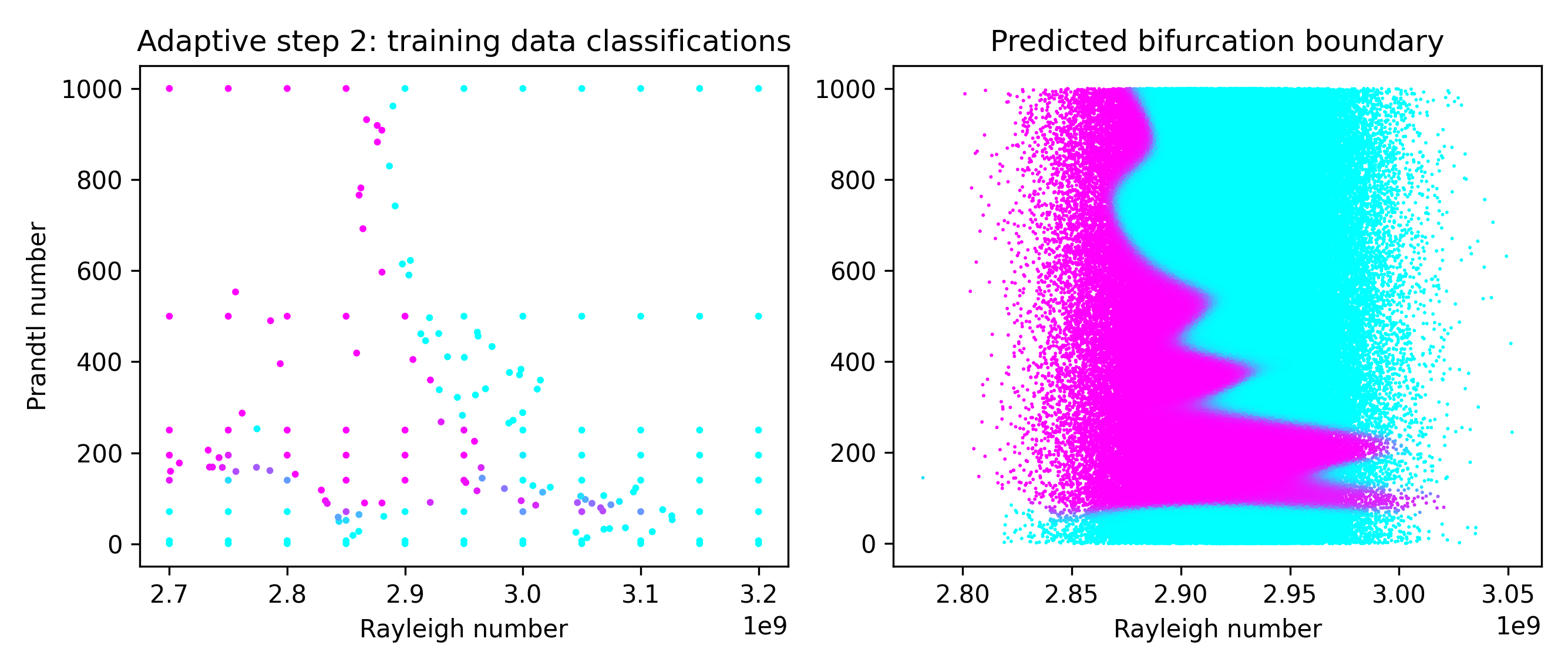}
	\includegraphics[width=0.95\linewidth]{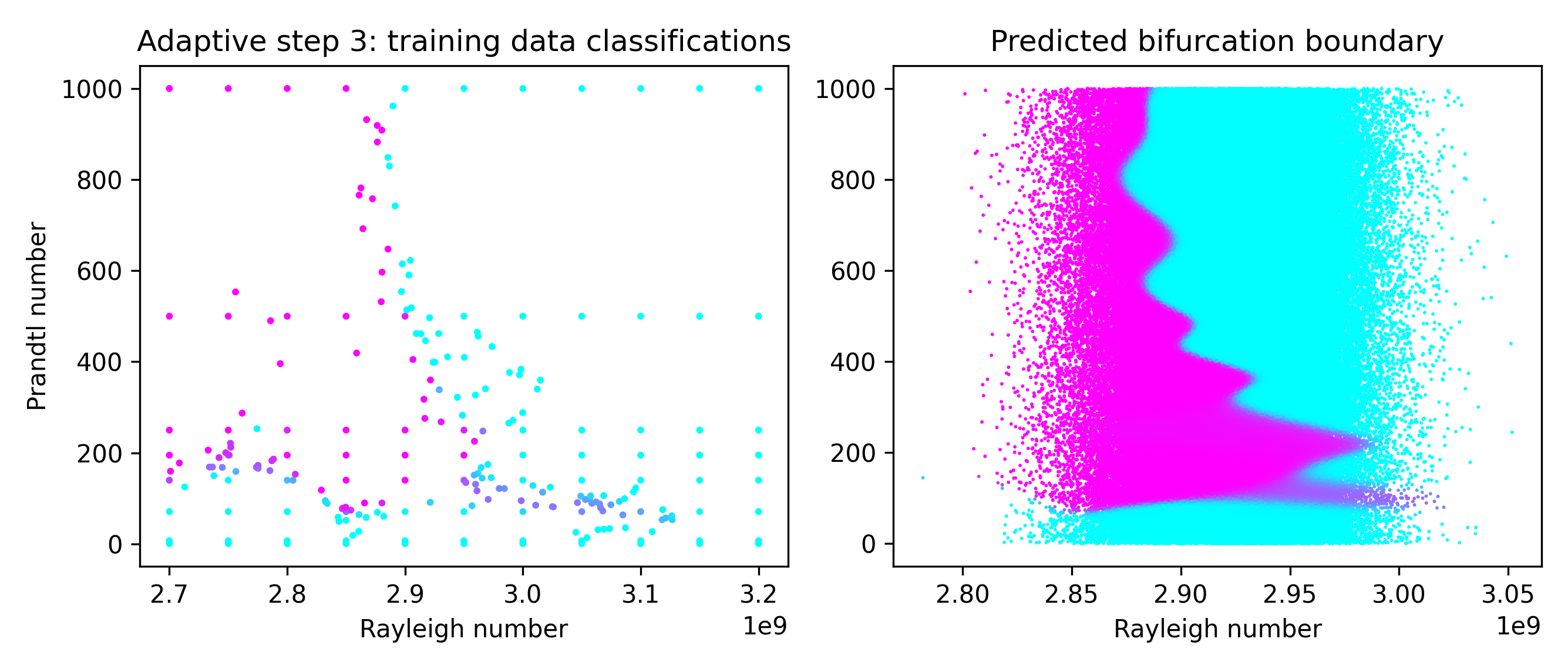}
	\caption{Adaptive refinement of the training set and the resulting bifurcation boundary for the differentially heated cavity problem. Each row shows the training data (left) and the classifier-predicted boundary (right) at the second and third adaptive refinement steps.}\label{problem3_boundary_plots_part2}
\end{figure}

Further, Figure~\ref{problem3_boundary_plots_part2} shows the outcome of second and third refinement steps. The second update has the most pronounced effect: KRnet again focuses its samples along the changing transition zone, creating a denser and more informative training set. In contrast, the third refinement makes only small adjustments, suggesting that the classifier-generator approach has nearly converged. Although all steps are displayed for completeness, the essential structure of the bifurcation boundary is already well resolved after the second adaptive step.

It is worth noting that the labelling process for this problem is significantly more expensive than in the previous examples. For each parameter pair $(\mathrm{Ra},\mathrm{Pr})$, the IFISS simulation typically requires four hours to determine whether the flow settles to a steady state or develops a periodic oscillation. As a result, obtaining labels through direct CFD solves is computationally expensive. The adaptive sampling strategy therefore plays a crucial role in reducing the total number of simulations required by concentrating new samples only in regions where the classifier is uncertain. This provides a practical advantage in high-cost settings such as the differentially heated cavity, where each individual simulation represents a substantial computational investment.

\vspace{0.2cm}
\noindent Across all three problems, the adaptive sampling strategy was able to learn the bifurcation boundaries using only a small set of initially uniform samples and without any prior assumptions about the structure of the parameter space. This contrasts with the earlier study \cite{silvester2026machine}, which depended on problem-specific heuristics to construct the training set. By reducing the number of expensive IFISS simulations needed to resolve the stability transition, the proposed approach offers a flexible and efficient alternative for exploring complex fluid-dynamical bifurcations.

\section{Conclusions}\label{sec:conclusion}

This study presents a novel adaptive, uncertainty-driven approach for mapping bifurcation
boundaries in fluid flows using a combination of neural-network classification and a
flow-based generative model. In contrast to previous approaches that rely on dense,
structured parameter sweeps guided by prior knowledge of the stability region, the
proposed method automatically concentrates CFD simulations in regions where they are
most informative. A key advantage of this approach is its ability to refine the training
set without any manual tuning of numerical perturbations or prescribing parameter samples in advance. Once coupled with the classifier, KRnet proposes new points based purely from predictive uncertainty, yielding sharp and reliable bifurcation boundaries even in regimes where the underlying physics is highly sensitive to parameter changes. In this way, the training set is enriched adaptively based on how the classifier behaves across the parameter domain. This mirrors the philosophy of adaptive mesh refinement in classical numerical methods: regions near rapidly changing segments of the bifurcation boundary naturally attract more samples, while areas that vary smoothly require far fewer evaluations. This targeted refinement explains the efficiency of the proposed strategy and its ability to resolve stability boundaries accurately while keeping the number of CFD simulations to a
minimum. The results for all three problems demonstrated that the method
yields smooth, physically meaningful classifications and converges rapidly to the correct
stability structure.

Although the present study focused on two-parameter problems in two-dimensional
settings, the methodology readily generalises to higher-dimensional parameter spaces
due to the scalable structure of KRnet and the entropy-based uncertainty quantification.

\section*{Reproducibility Statement}

All computational results reported in this paper are reproducible. 
The hydrodynamic stability datasets were generated using the IFISS software package 
(version~3.7), which is available from 
\url{https://www.manchester.ac.uk/ifiss} and 
\url{https://github.com/mcbssds/IFISS_download}. 
Users must obtain IFISS separately.
The adaptive sampling framework incorporates the flow-based deep generative model KRnet, 
obtained from the authors' public repository (\url{https://github.com/MJfadeaway/DAS}); this code is 
an external dependency and is not redistributed here.
Our MATLAB wrapper scripts for running IFISS in batch/parallel mode, together with the 
Python codes used for adaptive sampling, KR-NET integration, and surrogate model training, 
are available at \url{https://github.com/SinghA01-ml/adaptive-hydrostability}. 
Detailed installation and usage instructions are provided in the repository README.

\bibliographystyle{elsarticle-num}
\bibliography{mref}
\end{document}